# An improved kinetic Monte Carlo approach for epitaxial submonolayer growth

Robert Deák[1,2], Zoltán Néda[2] and Péter B. Barna[3, *]

[1]Eötvös Loránd University, Department of Materials Science, Budapest, Hungary
[2]Babeş-Bolyai University, Department of Theoretical and Computational Physics, Cluj-Napoca, RO-400084, Romania
[3]Research Institute for Technical Physics and Materials Science, H-1525, Budapest, P.O box 49, Hungary

**Abstract**

Two-component submonolayer growth on triangular lattice is qualitatively studied by kinetic Monte Carlo techniques. The hopping barrier governing surface diffusion of the atoms is estimated with an improved formula and using realistic pair interaction potentials. Realistic degrees of freedoms enhancing the surface diffusion of atoms are also introduced. The main advantages of the presented technique are the reduced number of free parameters and the clear diffusion activated mechanism for the segregation of different types of atoms. The potential of this method is exemplified by reproducing (i) vacancy and stacking fault related phase-boundary creation and dynamics; (ii) a special co-deposition and segregation process where the segregated atoms of the second component surrounds the islands formed by the first type of atoms.

**1. Introduction**

Simultaneous deposition of different types of atoms is widely encountered in experiments and practical applications [1-8]. It can be used for engineering special coating structures (e.g. nanocomposites by inhibitor additive) [9-13], or for improving the quality of the epitaxially grown films by using one component as a kind of surfactant [14,15]. On the other hand impurities (contaminants), operating generally as inhibitors, are always present during the deposition process, and even a minute amount of them can drastically modify both the bulk structure and the surface growth morphology of the films [16-20]. Typical topological features related to inhibitor impurity effects are: (i) irregular shapes of monolayer islands, (ii) bunches of growth steps forming hillocks and dents within the surface of crystals, (iii) truncated and rounded crystal shapes, as well as (iv) deep grain boundary grooves decorated by small crystals in polycrystalline films. Appearance of repeated nucleation and islands on the surface indicates directly that the crystal growth is interrupted by a surface covering layer i.e. the crystals became encapsulated by the impurity phase [1, 18, 19, 21, 22]. This phenomenon was clearly demonstrated by in situ transmission electron microscopy experiments in carbon contaminated indium films [1].

---

* corresponding author: Peter. B. Barna (barnap@mfa.kfki.hu)

The ideal one-component deposition is thus seldom realized and one always encounters the situation where species of several material components participate in the surface atomic processes. Foreign species can control the course of the fundamental phenomena and the pathway of structure evolution. To understand the effect of surfactant or inhibitor impurities on the structure evolution and the complex atomic processes taking place on the growth surface a consecrated method is to use kinetic Monte Carlo (MC) simulations [23-26]. This computational approach can help researchers in predicting the pathway of structure evolution and the structures that will form at different experimental conditions. By this way one could understand also the effect of the experimentally controllable parameters and engineer structures with desired practical properties.

It is well known that MC simulations are much weaker approximations to reality than the nowadays fashionable ab-initio Molecular Dynamics simulations [27]. By considering kinetic MC simulations several processes are taken into-account only in a phenomenological manner, without considering a microscopically realistic mechanism for it. The interaction potentials governing the dynamics of the atoms are also heuristic and usually they are not derived from first principles. MC simulations offer however a great advantage (for a review see [28]): it is fast and one can study thus larger systems and longer time-scales. Due to this advantage it is also more adaptable for moderate computational resources than ab-initio Molecular Dynamics. Quite reasonable number of atoms can be studied on cheep PC type computers. Making MC simulations more realistic is an important task. This could help researchers in elaborating powerful codes for predicting developing structure of thin-films or the topology of auto-organized nano-structures. Here we use a fast and microscopically more realistic kinetic Monte Carlo method for two-component sub-monolayer growths. The presented method can be generalized for several co-deposited components and also for the case of multilayer growth.

First the generally used kinetic Monte Carlo techniques are presented and the specific problem considered in the present work is discussed. Then the improved kinetic Monte Carlo approach is described and applied for the targeted problem.

**2. Kinetic Monte Carlo methods for epitaxial monolayer growth**

Kinetic (or resident time) Monte Carlo methods are appropriate for simulating those dynamical phenomena where several processes with widely different rates are simultaneously present. In the case of pattern formation during epitaxial growth this is the case: (i) atoms can be deposited on a crystalline surface with a given rate, (ii) atoms can diffuse on the surface (this diffusion being governed again by different rates as a function of the binding energy of the specific atom) and (iii) decohesion of adatoms from the surface are also possible (Fig. 1). When dealing with several material components the different type of neighbouring atoms can exchange by interchanging their positions (Fig. 1) following complicated microscopic mechanisms. This exchange-segregation process is characterized again with a different rate. Sometimes this is the key process for controlling the structure evolution and, consequently, for engineering specific (e..g. nanocomposite) structures in multi-component systems.

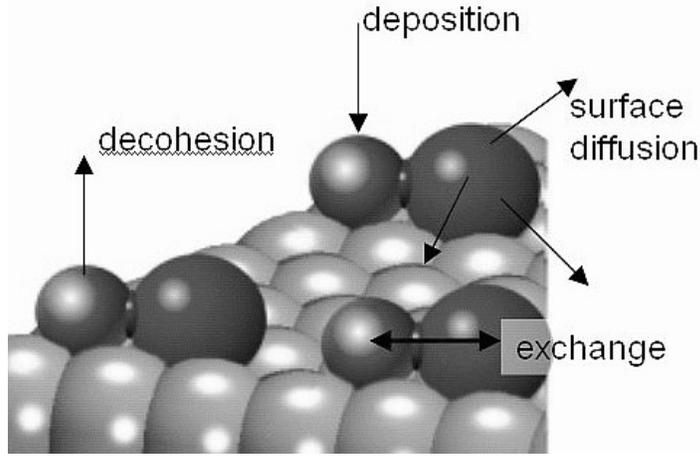

**Fig. 1** *Processes with widely different rates that govern the dynamics of atoms during epitaxial growth.*

In order to save precious computational time in simulating these processes with widely different rates, a Monte Carlo method used for studying equilibrium properties of low temperature systems (the BKL Monte Carlo method [23]) was adapted and named as kinetic Monte Carlo method (for a review see [28]). The basic idea is that in each simulation step one process is probabilistically selected (the probability to select one process is proportional with it's rate) and carried out. The time is than updated non-uniformly, depending on the rates of all possible processes at that given moment.

Generally the **deposition rate** is fixed and calculated from the deposition speed (deposition flux) given as the number of new monolayers deposited in unit time (ML/s).

The **diffusion rate** ($r_{X \to Y}$) of an atom is governed by the thermodynamic temperature ($T$) of the system and the potential barrier ($\Delta E_{X \to Y}$) that the atom has to overcome between the initial ($X$) and the final ($Y$) position:

$$r_{X \to Y} = f_0 \exp(-\frac{\Delta E_{X \to Y}}{kT}) \qquad (1)$$

In expression (1) $k$ is the Boltzmann factor and $f_0$ is the attempt rate, which is roughly the vibration frequency of atoms in the crystal ($f_0 \approx 10^{12}$ Hz). Since the value of the barrier is not straightforward to estimate (even if the pair-potential between the atoms is known), several simplifying methods are used. The simplest approach is to consider the potential barrier dependent only on the binding energy of the atom in the initial X state [29-32] or by applying the transition state theory [33]. A better, but computationally more costly approach is to consider a realistic pair-potential between the atoms [34] and estimate the potential in several points between the initial and final state. The embedded-atom method [35,36] offers another possibility for estimating the potential barrier in the hopping process. In such case the difference between the maximum and initial value will yield the potential barrier. An even more complex approach would be to map the potential in the neighbourhood of the initial and final point not on a line, but on a surface.

The **decohesion rate** is obtained either by fixing a potential barrier $E_{dec}$ for this process or by calculating the more realistic potential barrier as the binding energy of the chosen atom at the given site.

Exchange between neighbouring and different type of atoms are microscopically realized through complex vacancy mechanisms. In simulations however an oversimplified geometry is considered where many possible degrees of freedom for diffusion are not considered, so exchange possibilities are mostly blocked. The **exchange rate** is then usually postulated in form (1) by assigning a hypothetical $E_{ex}$ potential barrier for this process.

Simulations are usually performed in a two-dimensional geometry [29-32], the atoms being allowed to occupy only the sites of a pre-defined lattice. By this approach one reproduces an idealized situation where a new layer is growing on a perfect crystalline substrate. The simplest possibility is to consider a square lattice and the sites on the growing layer positioned exactly on the top of the atoms forming the substrate [29-32]. In such manner a non-realistic three-dimensional cubic structure is simulated but approaches on more complex geometries are also possible. One can use lattices with different symmetries and different stacking sequences for positioning the atoms in the growing layer [28]. Simulations can be made more realistic by considering a second layer on the top of the simulated one so that interchanges between these two layers become possible. This would allow formation of additional defects and vacancies.

Nowadays computationally costly off-lattice kinetic Monte Carlo methods [37, 38] are also considered for the case when several types of atoms are simultaneously present and there is lattice constant or symmetry mismatch between the crystalline structures of the components. In such an approach the position of the atoms are computed from an energy minimization procedure and the dynamics of the system is realized with the kinetic Monte Carlo algorithm. The method is an optimal reconciliation between the realistic nature of the Molecular Dynamics simulations and the higher speed of the kinetic Monte Carlo approach. Beside many interesting results obtained with this approach the method is still not usable for reasonably large system sizes in (2+1)D and practically relevant dynamics times.

**3. The considered problem and previous results**

A problem which was several times studied in the literature by kinetic Monte Carlo simulations is reconsidered and analyzed in the present work by an improved approach. Atoms of type **A** (growing materials) and **B** (impurities) are co-deposited on the planar surface of a perfect 3D single crystal of **A** type of atoms [31, 32]. There is a deposition rate for both components ($F_A$ and $F_B$ respectively, with $F_A=F_B$ for simplicity), and the atoms deposited on the substrate are allowed to diffuse there. The conditions necessary to obtain a particular segregation of the **A** and **B** atoms where the **B** impurities will surround

(decorate) the islands formed by the **A** adatoms are investigated. Earlier kinetic Monte Carlo simulations [31,32] performed on simple square-lattice geometry and using a first approximation for the $\Delta E_{X \to Y}$ hopping barrier concluded the important result that such structures can be obtained only if a direct exchange mechanism between **A** and **B** atoms on neighbouring sites is postulated.

More precisely in the earlier simulations it is considered that the hopping barrier for an atom of type W ($\Delta E^W_{X \to Y}$) depends only on the initial state of the atom. It is the sum of a term related to the substrate, $E_{sub}$, and a contribution related to each lateral nearest neighbour. Contributions depend on the local composition so that for each term we have four possibilities: **AA**, **AB**, **BA** and **BB**. The hopping barrier is then calculated as

$$\Delta E^W_{X \to Y} = E^W_h = \sum_{Q \equiv A,B} (n^W_0 E^{WQ}_{sub} + n^{WQ}_1 E^{WQ}_n), \tag{2}$$

where $n^W_0$ is $1$ if the substrate atom is of type $W$ and $0$ otherwise, $n^{WQ}_1$ is the number of nearest-neighbour W-Q pairs, $E^{WQ}_n$ is the corresponding contribution to the barrier (symmetric in W and Q, $E^{WQ}_n = E^{QW}_n$) and $E^{WQ}_{sub}$ is the contribution from a free $W$ atom on a substrate atom $Q$. In order to perform the kinetic Monte Carlo simulation for this system one has to postulate the $E^{WQ}_{sub}$ and $E^{WQ}_n$ values and the $f_0$ attempt frequency. The $T$ temperature of the system and the $F_A = F_B$ deposition rate has to be selected. Simulations were performed on a square lattice where the positions of the atoms in the growing layer are on the top of the substrate atoms.

These simulations proved that an energetic bias favouring segregation is not sufficient to obtain configurations with impurities (**B**) mostly positioned at island edges. To achieve this peculiar segregation a thermally activated exchange mechanism had to be introduced between **A** and **B** atoms on neighbouring sites. The easiest way to realize this was postulating a phenomenological potential barrier $E_{ex}$, for this process and to use equation (1) for calculating the exchange rate.

The present work intends to further improve the kinetic Monte Carlo simulation for the two-component system. Our task is to reduce considerably the number of postulated parameters (e.g. $E^{WQ}_{sub}$ and $E^{WQ}_n$) and to consider a more isotropic geometry which will increase the degrees of freedom for diffusion. As a consequence of this, for a reasonably large parameter set the interchange of **A** and **B** atoms will arise automatically and the phenomenological parameter $E_{ex}$ is eliminated. In several cases when the energetic bias favours segregation impurity decorated islands will form.

**4. The improved kinetic Monte Carlo approach**

In the present kinetic MC algorithm less parameters is postulated, diffusion of atoms with increased degrees of freedom and improved potential barriers are considered. At the same time the speed of the algorithm does not drop in considerable manner (the advantage of the MC approach is kept).

The first modification is that the triangular lattice ((111) plane of fcc structure) is used as substrate (filled circles in fig. 2). This leads also to a more compact packing of the atoms. It is assumed that atoms are spheres with the same diameter for both **A** and **B** components. In such manner there are two triangular sub-lattices (empty circles and crosses in fig. 2) on which the adatoms can be deposited forming monolayer lattices of fcc or hcp crystalline phases. Due to geometric restrictions atoms in the growing layer cannot occupy neighbouring sites belonging to different sub-lattices.

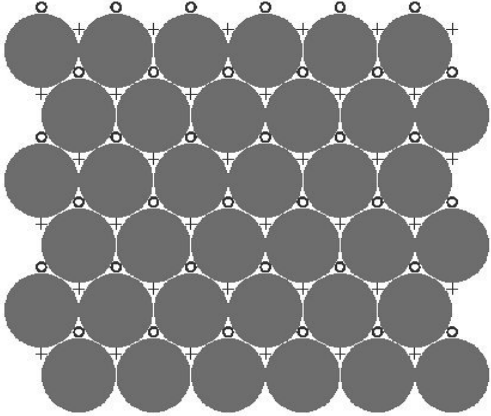

**Fig. 2** *Geometry of the considered lattice. Filled in large circles represent the atoms of the substrate, small empty circles and crosses represents the fcc and hcp lattice sites, respectively, on which the new layer can growth.*

Considering a bulk fcc substrate, stacking fault develops at the interface of the substrate and a growing hcp monolayer island (Fig. 3). By this way phase boundaries will also appear between growing islands of fcc and hcp types. This extra defect mechanism characteristic for this geometry facilitates the diffusion and interchange of atoms between the islands of the two growing phases. For the case of the two-component system only the growth of the fcc phase will be investigated, however the formation of phase boundaries and their motions in simple homoepitaxy will be also examined.

Diffusion of adatoms on the top of the first fcc type growing monolayer is also considered. These adatoms can also jump down on the substrate.

Apart of geometry a second improvement in the kinetic MC algorithm is in the calculation mode of the hopping barrier for the diffusion. First, realistic pair-potentials between the atoms are used to compute the binding energies of the atoms. The hopping barrier for the diffusion process is then calculated from the binding energies in the initial and final states.

A Lenard–Jones type of pair-potential was considered, although the program permits easily to change it to any other type of accepted form.

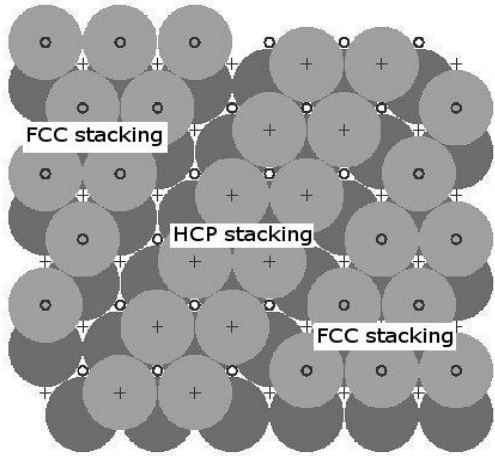

*Fig. 3.* The fcc and hcp phases that can form on a perfect fcc bulk substrate

Assuming the lattice constant as length unit, the interaction potential between atoms of type $W$ and $Q$ separated with distance $r$ can be written with one $E_{WQ}$ parameter:

$$U_{WQ} = E_{WQ}\left(\frac{1}{r^{12}} - \frac{2}{r^{6}}\right) \quad (3)$$

The parameter $E_{WQ}$ will fix the binding energy at the $r=1$ equilibrium distance. Interaction between atoms are taken into account up to $s=3$ lattice site distances. It is assumed that the hopping barrier from a site $X$ to a site $Y$ should depend not only on the binding energy in site $X$, but it should also depend on the change in the binding energy. The following form for calculating the hopping barrier of an atom has been proposed [39]:

$$\Delta E_{X \to Y} = -\alpha E_n^X + (1-\alpha)(E_n^Y - E_n^X) \quad (4)$$

In equation (4) $E_n^X$ ($E_n^Y$) is the total interaction energy (binding energy) of the atom at sites X (Y), respectively. α is a parameter between 0 and 1, whose value will be determined later. This is the simplest linear form in which the barrier depends both on the binding energy in the initial position and on the difference between the binding energies of the final and initial sites. Moreover, it yields the good barrier for decohesion ($-E_n^X$) and reasonable values for the self-surface diffusion and edge-diffusion. In order to get positive barriers for each possible process α has to be bounded between *0.3* and *0.6*. A simple exercise using Lenard-Jones type potentials on an fcc structure shows that the ratio of the energy barrier for self-surface diffusion and adsorption energy should be around *0.35*. It is immediate to realize that this ratio is exactly the value of α, and a first estimate is $\alpha = 0.35$.

The number of postulated parameters in the simulation code is strongly reduced. One has to fix only the values $E_{AA}, E_{AB}, E_{BB}$, the $F_A=F_B$ deposition rate, the $f_0$ attempt frequency and the *T* thermodynamic temperature of the system. Parameters very similar to the one used in previous studies [31, 32] were considered. We have fixed $f_0=10^{12}$ Hz, $E_{AA}$ =0.15eV, $E_{BB}$ =0.0001 eV, so that aggregation of **A** particles are favoured relative to the aggregation of **B** particles.  $E_{AB} = E_{BA}$ is varied in the 0.02 – 0.12eV interval, the thermodynamic temperature *T* was considered in the realistic 450 – 650K interval and the deposition rate $F_A=F_B$ in the 0.01 – 10 ML/s domain. Systems with lattice sizes up to 500x500 were easily simulated in a few days on normal PC type computers (Pentium 4, 3.4 GHz). As will be discussed in the next section, the peculiar segregation with the **B** impurity atoms decorating the islands formed by the **A** atoms can be obtained for a quite reasonable parameter range.

## 5. Simulation results

In the present paper we will present only some qualitative results for illustrating the applicability of the used method. Specific and practically important problems can be than studied using the described algorithm.

### 5.1 Evolution and annihilation of stacking-faults and phase-boundaries on an fcc (1,1,1) surface.

The case when **A** type of atoms are deposited with a low deposition rate ($F_A$=10ML/s) on a planar fcc (1,1,1) surface are simulated. In this case monolayer domains of the two equivalent orientations but with different, fcc and hcp, sequence are nucleated and grown. It worst mentioning however, that although the fcc and hcp sites are geometrically equivalent the binding energy is slightly different, the fcc sites being energetically more stable. Formation, motion and annihilation of stacking faults related phase-boundaries appear and can be nicely followed during simulation (fig. 4). Some movies are also given on the home-page dedicated to this study [40].

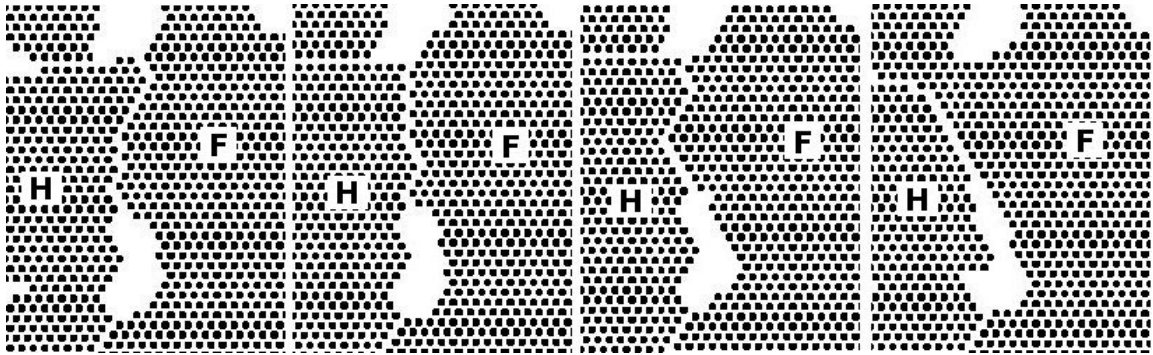

*Fig.4 Characteristic time evolution and annihilation of stacking faults related phase-boundaries for the case when only A type of atoms are deposited. The pictures from left to right represent some steps in the time-evolution. The F and H islands correspond to fcc and hcp stacking, respectively.  Simulation parameters are:  $E_{AA}$ =0.15eV, T=650K, $F_A$=10ML/s and $f_0=10^{12}$ Hz.*

### 5.2 Formation of impurity decorated islands in case of two-component deposition.

Simulations with the co-deposition of the **A** (•) and **B** (+) type of atoms leads to the expected structures. For the fixed parameters ($f_0=10^{12}$ Hz, $E_{AA}$ =0.15eV, $E_{BB}$ =0.0001 eV) and as a function of the $E_{AB}$ parameter two main type of structures are observable: (i) island containing intermixed **A** and **B** type of atoms, and islands decorated by **B** impurities. As expected, for low $E_{AB}$ values the impurity decorated islands are stable, while for higher $E_{AB}$ values the islands containing intermixed **A** and **B** type of atoms are observable (fig. 5). Increasing or decreasing the temperature will only shift the boundary between these two type of structures and favour larger or smaller islands of lower or higher number density (nucleation density) respectively, for the same number of deposited atoms.

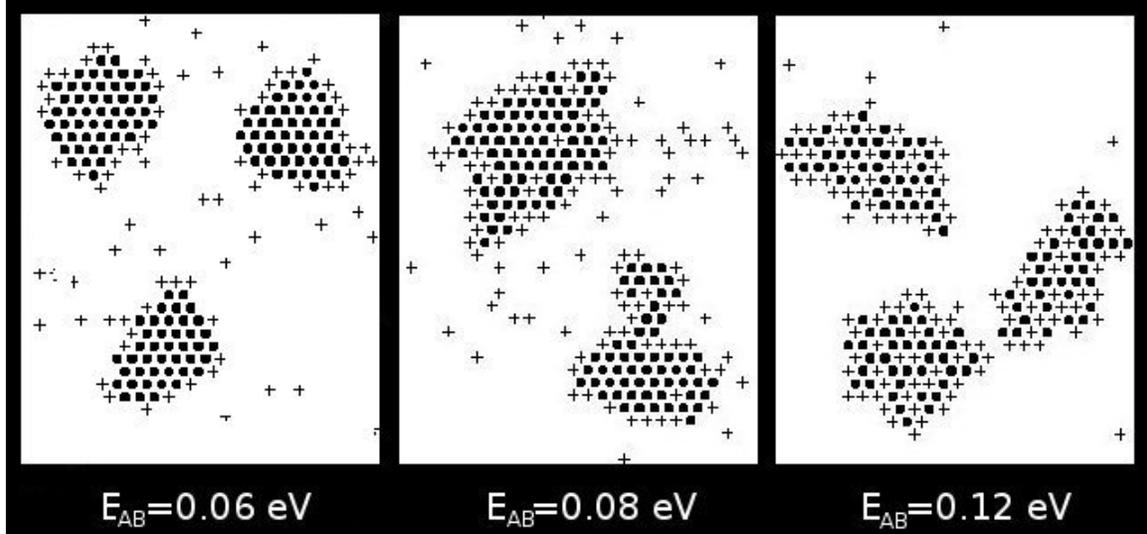

*Fig. 5 Island structures obtained as a function of the $E_{AB}$ parameter. Simulations with $E_{AA}$ =0.15eV, $E_{BB}$ =0.0001 eV, T=650K, $F_A=F_B$ =10ML/s and $f_0=10^{12}$ Hz. The structures from left to right are obtained for t=1,25 ·$10^{-2}$ s (60000 MCS), t=2,73 ·$10^{-2}$ s (20000 MCS) and t=1,87·$10^{-2}$ s (1500 MCS) simulation time, respectively. A central part of a much larger simulation area is presented. For $E_{AB}<0.08$ eV impurity decorated islands are formed.*

The time evolution of the structures for the case of impurity decorated islands is illustrated in Fig. 6. Some movies showing a more complete dynamics are available on the home-page [40] dedicated to this study.

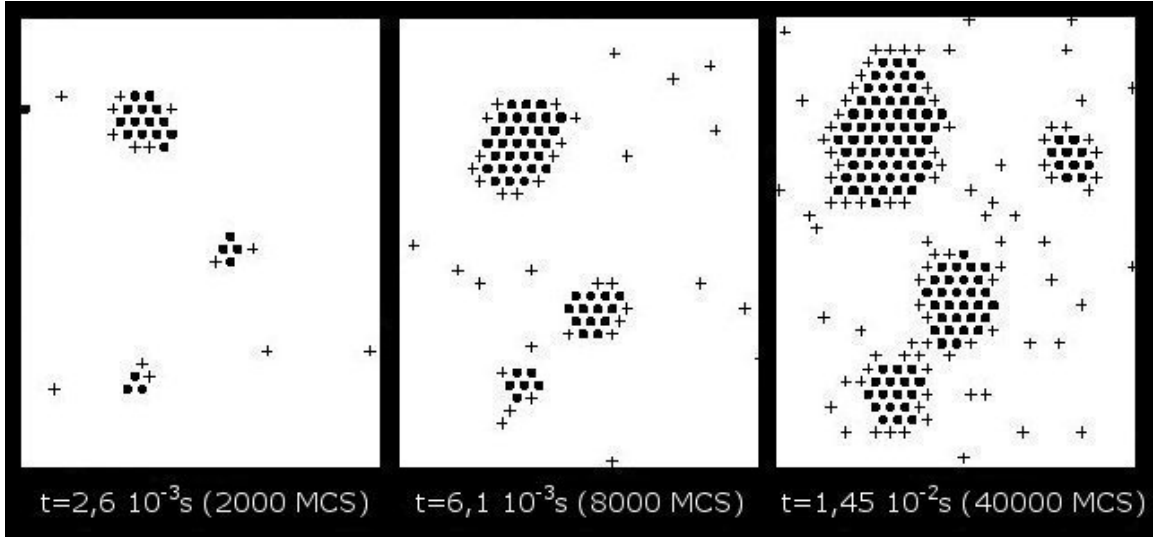

***Fig.6*** Snapshots from the time evolution of the system. Simulation parameters are $E_{AA}$ =0.15eV, $E_{BB}$ =0.0001 eV, $E_{AB}$=0.06 eV, T=650K, $F_A=F_B$ =10ML/s and $f_0=10^{12}$ Hz. A central part of a much larger simulation area is presented.

## 6. Discussion and Conclusions

The aim of the present work was to show that kinetic Monte Carlo methods can easily be improved without loosing it's main advantage of simulating large systems and reasonable long dynamics time. The two-component co-deposition process was successfully simulated by considering two main improvements relative to the simple classical method. First a new hopping barrier formula was used, calculated from the pair interaction potentials of the atoms (assumed to be of Lenard-Jones type in this study). Secondly, more degrees of freedom for the diffusion of particles were allowed by considering a triangular lattice and the growth of a second layer on the top of the simulated one. Interchanges of adatoms between these two layers are also possible. The exchange mechanism between atoms of different types on neighbouring lattice sites appears directly from diffusion without considering an artificially predefined rate for such a process. By this way it is possible to simulate with a reduced number of parameters the special segregation process in which impurities are decorating the growing islands. Vacancies, stacking faults, phase-boundaries and their dynamics are also successfully reproduced by this improved kinetic Monte Carlo technique. The method can be generalized to computationally study practically important pattern and structure formation problems during the co-deposition of several types of atoms on a given crystalline substrate. The presented method is computationally fast, and systems with tens of thousands of atoms can be simulated in reasonable computational time on PC type computers. The method is limited however to the case when the deposited components have similar crystal structures and similar lattice constants. Whenever there is lattice constant mismatch between the components off-lattice kinetic Monte Carlo methods have to be used [37, 38]. This will reduce however considerably the size of the system and the time length of the dynamics that can be investigated with a given computational resource.

**Acknowledgements.**

This work was supported by the Hungarian National Science Foundation under contract No. OTKA T048699, by INNOVATIAL 6[th] Framework Program Project IP 515844-21 and by Romanian CNCSIS grant 41/183. The authors acknowledge illuminating discussion with Prof. G. Radnoczi (Research Institute for Technical Physics and Materials Science, Budapest, Hungary).